\documentclass[letter,longauth,structabstract]{aa} 
\usepackage{graphicx}
\usepackage{txfonts}
\usepackage{natbib}
\bibpunct{(}{)}{;}{a}{}{,} 

\begin{document}
\title{HD 172555: detection of 63 $\rm \mu m$ [OI] emission in a
  debris disc\thanks{{\it Herschel} is an ESA space observatory with
    science instruments provided by European-led Principal
    Investigator consortia and with important participation from
    NASA.}}  
\author{P. Riviere-Marichalar\inst{1}, D. Barrado\inst{1,2}, J.-C. Augereau\inst{3},
  W. F. Thi\inst{3}, A. Roberge\inst{4},
  C. Eiroa\inst{5}, B. Montesinos\inst{1}, G. Meeus\inst{5},
  C. Howard\inst{6}, G. Sandell\inst{6},
  G. Duch\^{e}ne\inst{3,7},W. R. F. Dent\inst{8}, J. Lebreton\inst{3},
  I. Mendigut\'{i}a\inst{1,5}, N. Hu\'elamo\inst{1},
  F. M\'enard\inst{3,9}, C. Pinte\inst{3} }

   \institute{Centro de Astrobiolog\'{\i}a -- Depto. Astrof\'isica (CSIC--INTA), ESAC Campus, P.O. Box 78, 
	     28691 Villanueva de la Ca\~nada, Spain\\
             \email{riviere@cab.inta-csic.es}   
             \and Calar Alto Observatory, Centro Astron\'{o}mico Hispano-Alem\'{a}n C/Jes\'{u}s Durb\'{a}n Rem\'{o}n, 2-2, 04004 Almer\'{i}a, Spain 
             \and UJF-Grenoble 1 / CNRS-INSU, Institut de Plan\'{e}tologie et d'Astrophysique (IPAG) UMR 5274, Grenoble, F-38041, France 
             \and Exoplanets and Stellar Astrophysics Lab, NASA Goddard Space Flight Center, Code 667, Greenbelt, MD, 20771, USA 
             \and Dep. de F\'isica Te\'orica, Fac. de Ciencias, UAM Campus Cantoblanco, 28049 Madrid, Spain 
             \and SOFIA-USRA, NASA Ames Research Center 
             \and Astronomy Department, University of California, Berkeley CA 94720-3411 USA 
             \and ALMA, Avda Apoquindo 3846, Piso 19, Edificio Alsacia, Las Condes, Santiago, Chile 
             \and Laboratorio Franco-Chileno de Astronomia (UMI 3386: CNRS -- U de Chile / PUC / U Conception),  Santiago, Chile 
}
   \authorrunning{P.Riviere-Marichalar et al.}
   \date{}

 \abstract
{\object{HD~172555} is a young A7 star belonging to the $\beta$
  Pictoris Moving Group that harbours a debris disc. The
  \textit{Spitzer} IRS spectrum of the source showed mid-IR features
  such as silicates and glassy silica species, indicating the presence
  of a warm dust component with small grains, which places
  \object{HD~172555} among the small group of debris discs with such
  properties. The IRS spectrum also shows a possible emission of SiO
  gas.}
{We aim to study the dust distribution in the circumstellar disc of
  \object{HD~172555} and to asses the presence of gas in the debris
  disc.}
{ As part of the GASPS Open Time Key Programme, we obtained
  \textit{Herschel}-PACS photometric and spectroscopic observations of
  the source. We analysed PACS observations of \object{HD~172555} and
  modelled the Spectral Energy Distribution (SED) with a modified blackbody and the gas emission with
  a two-level population model with no collisional de-excitation.}
{We report for the first time the detection
  of [OI] atomic gas emission at 63.18 $\rm \mu m$ in the
  \object{HD~172555} circumstellar disc. We detect excesses due to
  circumstellar dust toward \object{HD~172555} in the three photometric
  bands of PACS (70, 100, and 160 $\rm \mu m$). We derive a large dust
  particle mass of $(4.8 \pm 0.6) \times 10^{-4}\, M_{\oplus}$ and an
  atomic oxygen mass of $2.5\times10^{-2}R^2$\,$ M_{\oplus}$, where
  $R$ in AU is the separation between the star and the inner disc. Thus, most of the detected mass of the disc is in the gaseous phase.}
{} \keywords{Stars: individual: HD 172555 -- Line: formation -- Kuiper
  belt: general -- Stars: pre-main sequence -- Circumstellar matter}
\maketitle
\section{Introduction}\label{Intro} 
Circumstellar discs play a key role in understanding how planets form
and evolve, since it is thought that they originate in gas-rich
circumstellar discs around young stars, the so-called protoplanetary
discs.  Around older stars, with ages greater than about 5 to 10~Myr,
another type of disc is seen: dusty, optically thin debris discs that
are produced by destructive collisions between planetesimals formed in
the protoplanetary disc phase. The younger debris discs ($\rm \sim$ 5
-- 100~Myr) are the likely sites of ongoing terrestrial planet
formation, while the older ones (0.1 -- 1~Gyr) correspond to the
``clearing out'' phase, when most planetesimals left over from planet
formation are removed from the system.

While debris dust has been extensively observed, previous studies have
provided little unambiguous information on the chemical composition
of the dust.  Most debris discs have featureless IR spectra, since the
grains are large and cold. However, there are a few debris discs that
show strong solid-state mid-IR spectral features, indicating the
presence of small, warm grains. These debris discs are probably highly
transient and are produced during periods of intense collisional grinding
between large bodies \citep{Lisse2008,Lisse2009}.  Gas can provide
additional compositional information, but it is rarely detected in
debris discs.  Only two -- \object{49~Ceti} and \object{HD~21997} --
show any trace of sub-mm CO emission, indicating that typical debris
gas abundances are low relative to those in younger protoplanetary
discs \citep{Dent2005,Moor2011}.  The only debris disc with a fairly
complete inventory of its gaseous species is $\beta$~Pictoris,
primarily obtained through UV/optical absorption spectroscopy
\citep{Roberge2006}.  In general, the debris gas studied so far
appears to be primarily low density and ionized, because
these low-density discs are optically thin to dissociating UV photons.

\object{HD~172555} \citep{Gray2006} is one of these rare stars with an observed warm debris disc.
This A7V-type star located at 29.2 pc \citep{Gray2006} belongs to the
$\beta$ Pictoris Moving Group (BPMG), and as such, its age is in the
range 12 Myr \citep{ZuckermanSong2004} to 20 Myr
\citep{Barrado1999}. \object{HD~172555} was originally found to
harbour a bright debris disc through IRAS measurements of excess
thermal emission attributed to circumstellar dust grains
\citep{Cote1987}. The {\it Spitzer}/IRS spectrum revealed strong solid-state 
emission features in the mid-IR, indicating that it
contains a relatively large amount of warm dust in a fine $\mu$m-sized
population at $T \sim 300$~K \citep{Lisse2009}. It has been proposed
that this disc is the aftermath of a catastrophic collision between
planetary mass bodies.

In this paper, we present \textit{Herschel} \citep{Pilbratt2010}
far-IR photometry and spectroscopy of \object{HD~172555}, obtained as
part of the ``Gas in Protopolanetary Systems'' (GASPS)
\textit{Herschel} Open Time Key Programme with the instrument PACS
\citep{Poglitsch2010}. In Sec.\,2 we describe the PACS spectroscopic
and photometric observations. In Sec.\,3 we describe the main
observational and modelling results for the dust and gas discs.  In
Sec.\,4, we discuss possible origins for the detected OI gas.

\section{Observations and data reduction}
\subsection{\textit{Herschel} photometry}
\begin{figure}[t]
\begin{center}
    \includegraphics[scale=0.5]{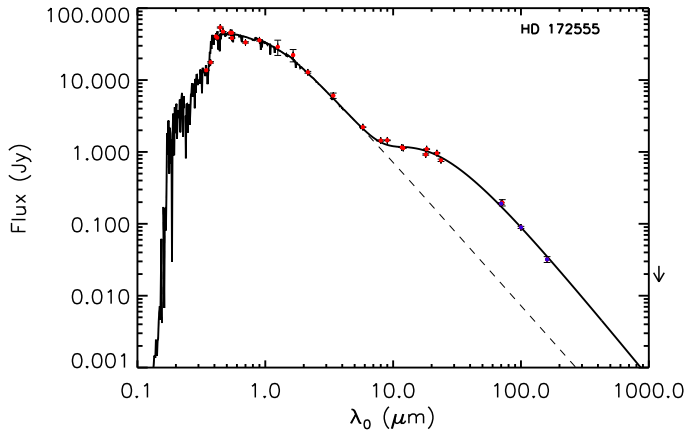}
   \caption{\object{HD~172555} SED. The dashed line represents the
     photospheric contribution fitted using VOSA \citep{Bayo2008} with
     $T_{\rm eff}=7800~$K, $\log g=4.5$, $ L_{\rm star}=7.8~L_{\sun}$,
     while the solid one represents a photosphere plus modified
     blackbody model with $T = 280$~K and $\rm \beta = 0.2$. Blue dots
     represent \textit{Heschel}-PACS photometric points.}
   \label{SED}
\end{center}
\end{figure}

\object{HD~172555} falls into the field of view of our scan maps
(obsids: 1342209059, 1342209060, 1342209061, 1342209062) centred on
\object{CD-64 1208}, an M0-type star with which it forms a wide binary
system on a $\rm \sim$ 2000 AU orbit \citep{Feigelson2006}. Photometry
was acquired with standard minimap configuration, with a scan length
of 3$\arcmin$ and scan angles 70$\degr$ and 110$\degr$ for the 70 and
100~$\rm \mu m$ bands. The \textit{Herschel}-PACS photometer observes
simultaneously in 70/100~$\rm \mu m$ and 160$\rm \mu m$, so this gave
us four maps at 160~$\rm \mu m$. Photometric observations were reduced using HIPE
8. We combined the final images at the same wavelength using IRAF
(http://iraf.noao.edu/) IMCOMBINE, weighting the images with the
exposure time. We performed aperture photometry with an aperture
radius of 6$\arcsec$ for the 70 and 100 $\rm \mu m$ bands and
12$\arcsec$ for the 160 $\rm \mu m$ band. Photometric errors were
computed by averaging the standard deviation in nine positions
surrounding the source scaled with the square root of the number of
pixels inside the aperture. Finally, we applied the aperture corrections
supplied by the PACS team. Photometric fluxes are listed in Table
\ref{HerschelPhot}. The source is point-like in the three PACS
bands. The PACS photometric points are shown in blue in
Fig. \ref{SED}.

\begin{table}
 \caption{\object{HD~172555} \textit{Herschel} Space Observatory
   photometry}
\label{HerschelPhot}
\centering
\begin{tabular}{l l l l l}
\hline \hline
Wavelength & Flux & Statistical & Calibration & Total \\
                    & density & error & error & error \\
($\rm \mu m$) & (mJy) & (mJy) & (mJy) & (mJy) \\
\hline
70 & 191 & 1.6 & 5.1 & 5.3  \\
100 & 89 & 1.8 & 2.4 & 3.0  \\
160 & 32 & 3.2 & 1.3  &  3.5  \\
\hline
\end{tabular}
\end{table}

\subsection{\textit{Herschel} spectroscopy}\label{Hspec}
\begin{figure}[t]
   \centering 
       \includegraphics[scale=0.48]{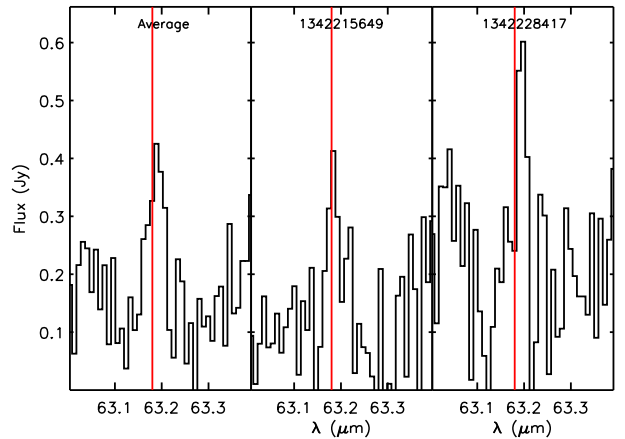}
      \caption{[OI] $\rm ^{3}P_{1} \rightarrow ^{3}P_{2}$ spectral
        feature. The vertical red lines show the positions in the rest
        frame wavelength for the [OI] $\rm^{3}P_{1}
        \rightarrow^{3}\!\!P_{2}$ transition at 63.18 $\rm \mu m$.}
   \label{OI63}
\end{figure}

The star was observed in chop/nod line spectroscopic mode with PACS on
March 8, 2011 and September 10, 2011. PACS allows imaging
of a $47\arcsec \times 47\arcsec$ field of view, resolved into
$5\times 5$ pixels known as spaxels. The observations were centred on
the 63.18 $\rm \mu m$ [OI] $\rm^{3}P_{1} \rightarrow^{3}\!\!P_{2}$
transition, within the wavelength range 62.94 to 64.44 $\rm \mu m$,
with a spectral resolution $\lambda / \Delta \lambda$ of $\rm
3570$. We also observed the star with PACS in pointed chopped range
spectroscopic mode on September 10, 2011, with the red channel centred
around 158 $\rm \mu m$, with a spectral resolution of $\rm 1265$. All
spectra were reduced using HIPE 7 and a modified version of the
reduction pipeline, including saturated and bad pixel removal, chop
subtraction, relative spectral response correction, and
flat-fielding. The spectra were extracted from the central spaxel and
aperture corrected to account for dispersion in the surrounding
spaxels. The final, reduced line spectra are shown in Fig. \ref{OI63}.
\section{Results}
\subsection{Star properties and evolutionary status}
A compilation of data from the literature was made to model
the Spectral Energy Distribution (SED, see Table \ref{Phot}). We also retrieved IRAC observations
from the \textit{Spitzer} Space Telescope archive (AOR 3923456 for
3.4, 4.5 and 5.0 $\rm \mu m$ bands and 25202432 for the 8.0 $\rm \mu
m$) and extracted aperture photometry using MOPEX. The IRAC 3.6 and
4.5 $\rm \mu m$ photometry is saturated.  The stellar parameters were
computed by fitting Johnson, Str\"omgren, and 2MASS photometry with
Kurucz photospheric models making use of the Virtual Observatory SED
Analyzer \citep[VOSA,][]{Bayo2008}. We find a value of $7.8 \pm
0.7~L_{\sun}$ for the stellar luminosity \citep[while ][finds $9.5~L_{\sun}$]{Wyatt2007} and $7800 \pm 200$~K
for the effective temperature, while \cite{Chen2006} finds $\rm T_{eff}=8550~K$, and \cite{Wyatt2007,Lisse2009} used $\rm T_{eff}=8000~K$.

\begin{figure}[t]
   \centering
     \includegraphics[scale=0.45, trim=0mm 0mm 0mm 0mm,clip]{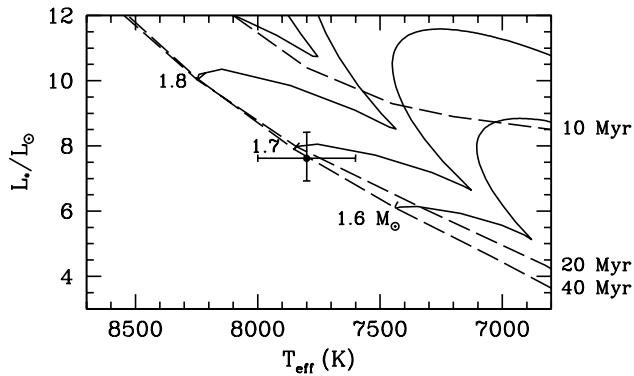}
   \caption{HR diagram for \object{HD~172555}. Dashed lines depict the
     position of the theoretical isochrones for different ages
     \citep{Yi2001} while solid lines depict evolutionary tracks for
     different stellar masses.}
   \label{HRD}
\end{figure}

Fig. \ref{HRD} shows the position of the star in the
Hertzsprung-Russell (HR) diagram. An error of $\pm 200$ K was
adopted for $T_{\rm eff} $ and the error bar in the luminosity
corresponds to the uncertainty of 0.2 pc. The
evolutionary tracks and isochrones from the Yonsei-Yale group
\citep{Yi2001} were used. Since the metallicity of this object is
almost solar ([M/H]=0.09, \cite{Gray2006}) the set with $Z\!=\!0.02$
(solar) was chosen. According to this comparison, we find $
M_{\star}\simeq 1.68~M_{\odot}$, and $\log
g_\star=4.28$, which implies $R_\star=1.56~R_\odot$. Given the
position of the star and the uncertainties, we can only set a lower
limit to the age of $\sim 17$~Myr (Fig. \ref{HRD}). The different ages quoted for BPMG
members range from 12 to 50~Myr, although the latter is an upper
limit.

\subsection{The dust disc}
\label{DustDisc}
As can be seen in Fig. \ref{SED}, little or no excess is detected at
wavelengths shorter than $7~\mu$m, while significant excess is seen at
longer wavelengths.

We used a modified blackbody model to fit the IR excess in
\object{HD~172555}: $F_{\rm cont}=B_{\nu} \times
(\lambda_{0}/\lambda)^{\beta}$ where $\rm B_{\nu}$ is the Planck
function, $\rm \lambda_{0}=13$\,$\rm \mu m$ and $\beta=0$ for
$\lambda < \lambda_0$. We do not use the $870~\mu$m LABOCA point
reported by \cite{Nilsson2009} because the association of the detected
2-$\rm \sigma$ signal with \object{HD~172555} is ambiguous.

The modified blackbody model gives a temperature of $280 \pm 9$~K,
with an opacity index (for $\lambda > \lambda_0$) $\beta = 0.2 \pm
0.2$, while a simple blackbody fit gives a higher temperature of $329
\pm 6$~K, but produces a poor fit to the SED. \cite{Rebull2008} fitted
the \object{HD~172555} IR excess with a blackbody at 310~K, in good
agreement with our determination. We note that we can fit the whole
SED with a single modified blackbody, implying that we do not see
evidence from this data set for a reservoir of colder dust producing
the far-IR emission. This means that the same dust population that
produces the mid-IR excess can also produce the entire
\textit{Herschel}/PACS excess.

Using $R_{\rm in} > \frac{1}{2} R_{*}\left(T_{*} \over T_{\rm
  dust}\right)^{(4+ \beta)/2}$ \citep{Beckwith1990}, we computed the
minimum radius for the dust distribution. We get $R_{\rm in,min}=4.0
\pm 0.3$~AU. The infrared excess computed by integrating the model is
$L_{\rm IR}/L_{*} \simeq 7.3 \times 10^{-4}$, slightly higher than the 
value shown in \cite{Wyatt2007} of $5 \times 10^{-4}$.  We used the flux at 160
$\rm \mu m$ to estimate the dust mass of the dust disc using $ M_{\rm
  dust}= F_{\nu}(160 \mu$m$)D^{2} / ( \kappa_{\nu} B_{\nu}(T_{\rm
  dust}))$, valid for optically thin discs, where $ F_{\nu}(160
\mu$m$)$ is the integrated flux density at 160 $\rm \mu m$, $D$ is the
distance to the star (29.2 pc), $\rm B_{\nu}$ can be approximated with
the Rayleigh-Jeans regime and $\rm \kappa_{\nu}=2\times (1.3 mm /
\lambda)^{\beta}~cm^{2}g^{-1}$. The final dust mass is $\rm (4.8 \pm
0.6) \times 10^{-4}$\,$ M_{\oplus}$.

\subsection{The gas disc}
\label{GasMass}
The [OI] $\rm^{3}P_{1} \rightarrow^{3}\!\!P_{2}$ transition at
63.18~$\mu$m was detected in both observations, with a signal-to-noise
ratio of $\rm \sim$4 in the average spectrum. The emission is only
detected on the central spaxel, indicating that it is centred on the
star. Line fluxes ($F_{\rm line}$) were obtained by fitting a Gaussian
plus continuum fit to the line and are shown in Table
\ref{lineFlux}. In both spectral line observations, the line is
spectrally unresolved within the uncertainties. The [CII] $\rm
^{2}P_{3/2} \rightarrow^{2}P_{1/2}$ emission line at 157.74~$\mu$m was
not detected; instead, we determine a 3-$\sigma$ upper limit of $2.3
\times 10^{-18}$\,$\rm W/m^{2}$ for its flux.
\begin{table}
\caption{\object{HD~172555} PACS line observations.}
\label{lineFlux}
\begin{tabular}{l l l l l l}
\hline \hline
Obs ID & Center & Width  & [OI] flux \\
-- & ($\rm \mu m$) & ($\rm \mu m$) & ($\rm 10^{-18} W/m^{2}$)  \\
\hline
1342215649 & 63.183 $\rm \pm$ 0.005 & 0.025 $\rm \pm$ 0.014 & $\rm 8.2 \pm 3.0$ \\
1342228417 & 63.195 $\rm \pm$ 0.005 & 0.018 $\rm \pm$ 0.005 & $\rm 9.1 \pm 3.8$ \\
Average spectrum &63.188 $\rm \pm$ 0.004 & 0.032 $\rm \pm$ 0.008 &$\rm 9.2 \pm 2.4$ \\
\hline
\end{tabular}
\end{table}

We computed the total [OI] mass using (see Appendix \ref{OxMass})
\begin{equation}\label{eqMass}
{\rm N_{[OI]}} \simeq 9.5\times 10^{36} \frac{F_{\rm line}({\rm
    erg\ s^{-1} \ cm^{-2}})\times D^2({\rm
    pc^2})}{\frac{h\nu}{4\pi}A_{ul}\ x(T) }\times R^2({\rm AU}^{2}),
\end{equation}
where ${\rm N_{[OI]}}$ is the number of oxygen atoms, $R$ is the inner radius,
 $A_{ul}$ is the Einstein $A$ coefficient, and $x$ is the
fractional population for the first excited state of the [OI]
$\rm^{3}P_{1} \rightarrow^{3}\!\!P_{2}$ transition. If the emission
originated at the minimal possible position of the dust ring at $\rm
\sim 4.0$~AU, the [OI] mass is $\rm \sim 0.39$\,$\rm
M_{\oplus}$. The computed mass is huge, so the reservoir for the [OI]
gas must have been very large. Nevertheless, the total atomic oxygen
mass strongly depends on the location where the emission comes from
(as $R^2$). For instance, the [OI] gas mass decreases to between a few
$10^{-4}$\,$\rm M_{\oplus}$ and a few $10^{-3}$\,$\rm M_{\oplus}$ if
it is located near the size-, and composition-dependent dust
sublimation radius ($\sim$0.1--0.5~AU). But we note that, even in this
case, the [OI] gas mass remains large compared to the dust mass.

\section{Origin of the gas}\label{Origin}

Our {\it Herschel}/PACS observations reveal a copious amount of atomic
oxygen, although the exact mass depends somewhat on its unconstrained
radial location (Sec.\,\ref{GasMass}). This raises the question of its
origin. \object{HD~172555} is a rare debris disc. The dust is located
in the inner system at only a few AU from the star
(Sec.\,\ref{DustDisc}), in contrast to typical debris discs composed
mainly of cold Kuiper belt-like dust at much larger distances from the star
\citep[e.g. the co-eval BPMG member
  \object{HD~181327},][]{Lebreton2012}. Here, we speculate about two
different origins for the observed [OI] gas.

\subsection{Recent release during a catastrophic event}
The models of \cite{Wyatt2007} argue that \object{HD~172555} infrared
fractional luminosity is $\rm \sim\!86$ times higher than the maximum
value expected from steady-state collisional evolution under their
modelling assumptions at an age of 12~Myr. Although a different
parameter selection could perhaps explain the departure from the
maximum value, the authors pointed out that a major stochastic event may be
playing a role.

Emission features in the \textit{Spitzer}/IRS spectrum of
\object{HD~172555} were modelled by \cite{Lisse2009}, who suggested
the presence of glassy silica. They also tentatively attributed faint
features to rovibrational transitions of SiO gas. They concluded that
the dust composition, and the presence of SiO gas (if confirmed),
point to a violent production process, such as an hypervelocity
collision between massive planetary bodies. This suggestion was
additionally supported by the hypervelocity collision experiments conducted
by \cite{Takasawa2011}. These authors found that the ejecta typically had a
steeper grain size distribution than expected for an equilibrium
distribution in a standard collisional cascade, as proposed by
\cite{Lisse2009} to describe the warm dust in \object{HD~172555} ($\rm dn/da \sim a^{-4}$). This
would argue for a source of fresh material, which would have had to be 
produced within the last 0.1~Myr to be consistent with SiO recondensation. Recently,
\cite{Pahlevan2011} showed that silicate constituents (Si, Fe, Mg and
O) can be produced in the form of gas during hypervelocity collisions.
Therefore, one can speculate that the origin of atomic oxygen detected
with {\it Herschel} could be the violent event proposed by
\cite{Lisse2009} to explain the shape of the \textit{Spitzer}-IRS
spectrum.

\subsection{Gas accumulation over time}

The extremely low sensivity of OI to the radiation pressure from
this A-type star \citep{Fernandez2006} opens the door to alternative,
perhaps less catastrophic scenarios as for the origin of the detected
OI gas. Atomic oxygen could indeed accumulate over time at the place
where it is produced. For instance, the considerable collisional activity
in the dust ring presumably releases a population of poorly bound, or
unbound grains due to radiation pressure, which can collisionaly
vaporize larger grains due to high relative velocities. This has been
proposed by \citet{Czechowski2007} to explain the stable gas in the
co-eval \object{$\beta$~Pictoris} system. Assuming that oxygen is
originally locked into olivine grains, a dust mass of $\sim
0.9~M_{\oplus}$ would need to have been vaporized in less than 12~Myr,
corresponding to a mass rate higher than $7\times
10^{-8}~M_{\oplus}/$yr ($10^{13}~$g/s). Interestingly, this value
compares to the mass rate derived by \citet{Czechowski2007} at the
peak position of the \object{$\beta$~Pictoris} dust disc (50--100~AU).

The sublimation of grains close to the star, at $\sim$0.1--0.5\,AU,
could also contribute to the production of OI. Although the
4~AU-radius dust ring is collision-dominated, models by \citet{Wyatt2005} and \citet{Loehne2012} show
that, even in this case, a fraction of the dust grains can escape the
collisional cascade and migrate toward the star by Poynting-Robertson
(P-R) drag. Silicates extracted from the ring will eventually reach
sufficiently high temperatures as they approach the star and produce
O, SiO and MgO, and ultimately Si, Mg and O. Assuming again that O is
locked in olivine grains, a reference dust mass of $\rm \sim
10^{-3}$\,$\rm M_{\oplus}$ would need to have been extracted from the
dust belt by P-R drag to explain the PACS data over an unknown time
span. A strict minimum supplying mass rate of about $\rm
10^{-10}$\,$\rm M_{\oplus}/yr$ is obtained assuming a unit efficiency
in atomic O production by sublimation over the longest possible time
span (12~Myr), and assuming that the gas has not viscously flown inward
or outward. On the other hand, the P-R drag timescale for bound grains
in the 4-AU radius dust ring, $t_{\rm PR}>7500$~yr
\citep[e.g.][]{Wyatt2005}, yields an upper limit on the present-day
P-R drag mass loss rate of a few $\rm 10^{-8}$\,$M_{\oplus}/$yr. This
implies that this process had to last for longer than a few $\rm 10^{4} yr$.

\subsection{Discussion}
Different scenarios can qualitatively explain the OI enrichment of
the \object{HD~172555} disc. We note that these are not necessarily
mutually exclusive since gas accumulation could follow the giant
collision proposed by \cite{Lisse2009}, if it did not occur too
recently. For instance, it takes about 1~Myr for the debris created in
a giant impact to form an axisymmetric disc \citep{Jackson2012}, and
be consistent with observational constraints \citep{Smith2012}. This
may leave sufficient time to accumulate OI gas at some place in the
disc. Additional processes, such as comet evaporation or grain
photodesorption \citep{Chen2007}, also deserve to be mentioned as
possible OI providers. In summary, the origin of the detected OI
gas remains open, and detailed modelling of the processes discussed
here, as well as the detection of atomic or ionized Fe, Mg, and Si 
will help to identify the source of OI gas.

\section{Summary and conclusion}
Our main results
can be summarized as follows:
\begin{enumerate}
\item We report the detection of the [OI] line in the debris disc of
  \object{HD~172555} with a flux of $\rm (9.2 \pm 2.4) \times
  10^{-18} W/m^{2}$. This is the first unambiguous detection of gas in
  this system.
\item We modelled the SED of the object with a Kurucz photospheric
  model ($T_{\rm eff}=7800~$K, $\log g=4.5$, $ L_{\rm star}=7.8~L_{\sun}$)
  plus a modified blackbody to fit the infrared excess at $
  \lambda > 7~\mu$m. We obtain $T_{\rm dust}=280~$K and $\beta
  =0.2$. We derive a dust mass of $\sim 4.8 \times 10^{-4}
  M_{\oplus}$.
\item Using purely analytical relations and simple assumptions, we
  derive an [OI] mass of $2.5\times10^{-2}R^2$\,$ M_{\oplus}$ where
  $R$ is the unconstrained radial location of the gas.
\item Although the spatial origin of the line is uncertain, we
  speculate about a possible origin in an hypervelocity collision, or
  in gas accumulation over time due to the low radiation pressure on
  atomic oxygen.
\item \object{HD~172555} provides a valuable window on the processes
  that occur during the early stages of terrestrial planet formation,
  including massive collisions like the one that produced Earth's moon
  \citep{Canup1996}.
\end{enumerate}

\acknowledgements This research has been funded by Spanish grants AYA
2010-21161-C02-02, CDS2006-00070 and
PRICIT-S2009/ESP-1496. J.-C. Augereau and J. Lebreton thank the ANR
(contract ANR-2010 BLAN-0505-01, EXOZODI) and the CNES-PNP for
financial support. C. Pinte, F. Menard and W.-F. Thi acknowledges funding from the EU
  FP7-2011 under Grant Agreement nr. 284405. G. Meeus is supported by RYC- 2011-07920. G. Meeus,
C. Eiroa, I. Mendigut\'{i}a and B. Montesinos are partly supported by AYA-2011-26202. FM acknowledges support from the Millennium Science Initiative (Chilean Ministry of Economy), through grant ÒNucleus P10-022-FÓ.

\bibliographystyle{aa} 
\bibliography{biblio.bib}

\begin{thebibliography}{37}
\expandafter\ifx\csname natexlab\endcsname\relax\def\natexlab#1{#1}\fi

\bibitem[{{Barrado y Navascu{\'e}s} {et~al.}(1999){Barrado y Navascu{\'e}s},
  {Stauffer}, {Song}, \& {Caillault}}]{Barrado1999}
{Barrado y Navascu{\'e}s}, D., {Stauffer}, J.~R., {Song}, I., \& {Caillault},
  J.-P. 1999, \apjl, 520, L123

\bibitem[{{Bayo} {et~al.}(2008){Bayo}, {Rodrigo}, {Barrado Y Navascu{\'e}s},
  {Solano}, {Guti{\'e}rrez}, {Morales-Calder{\'o}n}, \& {Allard}}]{Bayo2008}
{Bayo}, A., {Rodrigo}, C., {Barrado Y Navascu{\'e}s}, D., {et~al.} 2008, \aap,
  492, 277

\bibitem[{{Beckwith} {et~al.}(1990){Beckwith}, {Sargent}, {Chini}, \&
  {Guesten}}]{Beckwith1990}
{Beckwith}, S.~V.~W., {Sargent}, A.~I., {Chini}, R.~S., \& {Guesten}, R. 1990,
  \aj, 99, 924

\bibitem[{{Canup} \& {Esposito}(1996)}]{Canup1996}
{Canup}, R.~M. \& {Esposito}, L.~W. 1996, \icarus, 119, 427

\bibitem[{{Chen} {et~al.}(2007){Chen}, {Li}, {Bohac}, {Kim}, {Watson}, {van
  Cleve}, {Houck}, {Stapelfeldt}, {Werner}, {Rieke}, {Su}, {Marengo},
  {Backman}, {Beichman}, \& {Fazio}}]{Chen2007}
{Chen}, C.~H., {Li}, A., {Bohac}, C., {et~al.} 2007, \apj, 666, 466

\bibitem[{{Chen} {et~al.}(2006){Chen}, {Sargent}, {Bohac}, {Kim},
  {Leibensperger}, {Jura}, {Najita}, {Forrest}, {Watson}, {Sloan}, \&
  {Keller}}]{Chen2006}
{Chen}, C.~H., {Sargent}, B.~A., {Bohac}, C., {et~al.} 2006, \apjs, 166, 351

\bibitem[{{Cote}(1987)}]{Cote1987}
{Cote}, J. 1987, \aap, 181, 77

\bibitem[{{Czechowski} \& {Mann}(2007)}]{Czechowski2007}
{Czechowski}, A. \& {Mann}, I. 2007, \apj, 660, 1541

\bibitem[{{Dent} {et~al.}(2005){Dent}, {Greaves}, \& {Coulson}}]{Dent2005}
{Dent}, W.~R.~F., {Greaves}, J.~S., \& {Coulson}, I.~M. 2005, \mnras, 359, 663

\bibitem[{{Feigelson} {et~al.}(2006){Feigelson}, {Lawson}, {Stark}, {Townsley},
  \& {Garmire}}]{Feigelson2006}
{Feigelson}, E.~D., {Lawson}, W.~A., {Stark}, M., {Townsley}, L., \& {Garmire},
  G.~P. 2006, \aj, 131, 1730

\bibitem[{{Fern{\'a}ndez} {et~al.}(2006){Fern{\'a}ndez}, {Brandeker}, \&
  {Wu}}]{Fernandez2006}
{Fern{\'a}ndez}, R., {Brandeker}, A., \& {Wu}, Y. 2006, \apj, 643, 509

\bibitem[{{Gray} {et~al.}(2006){Gray}, {Corbally}, {Garrison}, {McFadden},
  {Bubar}, {McGahee}, {O'Donoghue}, \& {Knox}}]{Gray2006}
{Gray}, R.~O., {Corbally}, C.~J., {Garrison}, R.~F., {et~al.} 2006, \aj, 132,
  161

\bibitem[{{Hauck} \& {Mermilliod}(1998)}]{Hauck1998}
{Hauck}, B. \& {Mermilliod}, M. 1998, \aaps, 129, 431

\bibitem[{{H{\o}g} {et~al.}(2000){H{\o}g}, {Fabricius}, {Makarov}, {Urban},
  {Corbin}, {Wycoff}, {Bastian}, {Schwekendiek}, \& {Wicenec}}]{Hog2000}
{H{\o}g}, E., {Fabricius}, C., {Makarov}, V.~V., {et~al.} 2000, \aap, 355, L27

\bibitem[{{Ishihara} {et~al.}(2010){Ishihara}, {Onaka}, {Kataza}, {Salama},
  {Alfageme}, {Cassatella}, {Cox}, {Garc{\'{\i}}a-Lario}, {Stephenson},
  {Cohen}, {Fujishiro}, {Fujiwara}, {Hasegawa}, {Ita}, {Kim}, {Matsuhara},
  {Murakami}, {M{\"u}ller}, {Nakagawa}, {Ohyama}, {Oyabu}, {Pyo}, {Sakon},
  {Shibai}, {Takita}, {Tanab{\'e}}, {Uemizu}, {Ueno}, {Usui}, {Wada},
  {Watarai}, {Yamamura}, \& {Yamauchi}}]{Ishihara2010}
{Ishihara}, D., {Onaka}, T., {Kataza}, H., {et~al.} 2010, \aap, 514, A1+

\bibitem[{{Jackson} \& {Wyatt}(2012)}]{Jackson2012}
{Jackson}, A.~P. \& {Wyatt}, M.~C. 2012, \mnras, 425, 657

\bibitem[{{Johnson} {et~al.}(1966){Johnson}, {Mitchell}, {Iriarte}, \&
  {Wisniewski}}]{Johson1966}
{Johnson}, H.~L., {Mitchell}, R.~I., {Iriarte}, B., \& {Wisniewski}, W.~Z.
  1966, Communications of the Lunar and Planetary Laboratory, 4, 99

\bibitem[{{Lebreton} {et~al.}(2012){Lebreton}, {Augereau}, {Thi}, {Roberge},
  {Donaldson}, {Schneider}, {Maddison}, {M{\'e}nard}, {Riviere-Marichalar},
  {Mathews}, {Kamp}, {Pinte}, {Dent}, {Barrado}, {Duch{\^e}ne}, {Gonzalez},
  {Grady}, {Meeus}, {Pantin}, {Williams}, \& {Woitke}}]{Lebreton2012}
{Lebreton}, J., {Augereau}, J.-C., {Thi}, W.-F., {et~al.} 2012, \aap, 539, A17

\bibitem[{{Lisse} {et~al.}(2008){Lisse}, {Chen}, {Wyatt}, \&
  {Morlok}}]{Lisse2008}
{Lisse}, C.~M., {Chen}, C.~H., {Wyatt}, M.~C., \& {Morlok}, A. 2008, in Lunar
  and Planetary Inst. Technical Report, Vol.~39, Lunar and Planetary Institute
  Science Conference Abstracts, 2119

\bibitem[{{Lisse} {et~al.}(2009){Lisse}, {Chen}, {Wyatt}, {Morlok}, {Song},
  {Bryden}, \& {Sheehan}}]{Lisse2009}
{Lisse}, C.~M., {Chen}, C.~H., {Wyatt}, M.~C., {et~al.} 2009, \apj, 701, 2019

\bibitem[{{L{\"o}hne} {et~al.}(2012){L{\"o}hne}, {Augereau}, {Ertel},
  {Marshall}, {Eiroa}, {Mora}, {Absil}, {Stapelfeldt}, {Th{\'e}bault}, {Bayo},
  {Del Burgo}, {Danchi}, {Krivov}, {Lebreton}, {Letawe}, {Magain}, {Maldonado},
  {Montesinos}, {Pilbratt}, {White}, \& {Wolf}}]{Loehne2012}
{L{\"o}hne}, T., {Augereau}, J.-C., {Ertel}, S., {et~al.} 2012, \aap, 537, A110

\bibitem[{{Moerchen} {et~al.}(2010){Moerchen}, {Telesco}, \&
  {Packham}}]{Moerchen2010}
{Moerchen}, M.~M., {Telesco}, C.~M., \& {Packham}, C. 2010, \apj, 723, 1418

\bibitem[{{Mo{\'o}r} {et~al.}(2011){Mo{\'o}r}, {{\'A}brah{\'a}m}, {Juh{\'a}sz},
  {Kiss}, {Pascucci}, {K{\'o}sp{\'a}l}, {Apai}, {Henning}, {Csengeri}, \&
  {Grady}}]{Moor2011}
{Mo{\'o}r}, A., {{\'A}brah{\'a}m}, P., {Juh{\'a}sz}, A., {et~al.} 2011, \apjl,
  740, L7

\bibitem[{{Nilsson} {et~al.}(2009){Nilsson}, {Liseau}, {Brandeker}, {Olofsson},
  {Risacher}, {Fridlund}, \& {Pilbratt}}]{Nilsson2009}
{Nilsson}, R., {Liseau}, R., {Brandeker}, A., {et~al.} 2009, \aap, 508, 1057

\bibitem[{{Pahlevan} {et~al.}(2011){Pahlevan}, {Stevenson}, \&
  {Eiler}}]{Pahlevan2011}
{Pahlevan}, K., {Stevenson}, D.~J., \& {Eiler}, J.~M. 2011, Earth and Planetary
  Science Letters, 301, 433

\bibitem[{{Pilbratt} {et~al.}(2010){Pilbratt}, {Riedinger}, {Passvogel},
  {Crone}, {Doyle}, {Gageur}, {Heras}, {Jewell}, {Metcalfe}, {Ott}, \&
  {Schmidt}}]{Pilbratt2010}
{Pilbratt}, G.~L., {Riedinger}, J.~R., {Passvogel}, T., {et~al.} 2010, \aap,
  518, L1+

\bibitem[{{Poglitsch} {et~al.}(2010){Poglitsch}, {Waelkens}, {Geis},
  {Feuchtgruber}, {Vandenbussche}, {Rodriguez}, {Krause}, {Renotte}, {van
  Hoof}, {Saraceno}, {Cepa}, {Kerschbaum}, {Agn{\`e}se}, {Ali}, {Altieri},
  {Andreani}, {Augueres}, {Balog}, {Barl}, {Bauer}, {Belbachir}, {Benedettini},
  {Billot}, {Boulade}, {Bischof}, {Blommaert}, {Callut}, {Cara}, {Cerulli},
  {Cesarsky}, {Contursi}, {Creten}, {De Meester}, {Doublier}, {Doumayrou},
  {Duband}, {Exter}, {Genzel}, {Gillis}, {Gr{\"o}zinger}, {Henning},
  {Herreros}, {Huygen}, {Inguscio}, {Jakob}, {Jamar}, {Jean}, {de Jong},
  {Katterloher}, {Kiss}, {Klaas}, {Lemke}, {Lutz}, {Madden}, {Marquet},
  {Martignac}, {Mazy}, {Merken}, {Montfort}, {Morbidelli}, {M{\"u}ller},
  {Nielbock}, {Okumura}, {Orfei}, {Ottensamer}, {Pezzuto}, {Popesso},
  {Putzeys}, {Regibo}, {Reveret}, {Royer}, {Sauvage}, {Schreiber}, {Stegmaier},
  {Schmitt}, {Schubert}, {Sturm}, {Thiel}, {Tofani}, {Vavrek}, {Wetzstein},
  {Wieprecht}, \& {Wiezorrek}}]{Poglitsch2010}
{Poglitsch}, A., {Waelkens}, C., {Geis}, N., {et~al.} 2010, \aap, 518, L2+

\bibitem[{{Rebull} {et~al.}(2008){Rebull}, {Stapelfeldt}, {Werner}, {Mannings},
  {Chen}, {Stauffer}, {Smith}, {Song}, {Hines}, \& {Low}}]{Rebull2008}
{Rebull}, L.~M., {Stapelfeldt}, K.~R., {Werner}, M.~W., {et~al.} 2008, \apj,
  681, 1484

\bibitem[{{Roberge} {et~al.}(2006){Roberge}, {Feldman}, {Weinberger},
  {Deleuil}, \& {Bouret}}]{Roberge2006}
{Roberge}, A., {Feldman}, P.~D., {Weinberger}, A.~J., {Deleuil}, M., \&
  {Bouret}, J.-C. 2006, \nat, 441, 724

\bibitem[{{Sch{\"u}tz} {et~al.}(2005){Sch{\"u}tz}, {Meeus}, \&
  {Sterzik}}]{Schutz2005}
{Sch{\"u}tz}, O., {Meeus}, G., \& {Sterzik}, M.~F. 2005, \aap, 431, 175

\bibitem[{{Skrutskie} {et~al.}(2006){Skrutskie}, {Cutri}, {Stiening},
  {Weinberg}, {Schneider}, {Carpenter}, {Beichman}, {Capps}, {Chester},
  {Elias}, {Huchra}, {Liebert}, {Lonsdale}, {Monet}, {Price}, {Seitzer},
  {Jarrett}, {Kirkpatrick}, {Gizis}, {Howard}, {Evans}, {Fowler}, {Fullmer},
  {Hurt}, {Light}, {Kopan}, {Marsh}, {McCallon}, {Tam}, {Van Dyk}, \&
  {Wheelock}}]{Skrutskie2006}
{Skrutskie}, M.~F., {Cutri}, R.~M., {Stiening}, R., {et~al.} 2006, \aj, 131,
  1163

\bibitem[{{Smith} {et~al.}(2012){Smith}, {Wyatt}, \& {Haniff}}]{Smith2012}
{Smith}, R., {Wyatt}, M.~C., \& {Haniff}, C.~A. 2012, \mnras, 422, 2560

\bibitem[{{Takasawa} {et~al.}(2011){Takasawa}, {Nakamura}, {Kadono}, {Arakawa},
  {Dohi}, {Ohno}, {Seto}, {Maeda}, {Shigemori}, {Hironaka}, {Sakaiya},
  {Fujioka}, {Sano}, {Otani}, {Watari}, {Sangen}, {Setoh}, {Machii}, \&
  {Takeuchi}}]{Takasawa2011}
{Takasawa}, S., {Nakamura}, A.~M., {Kadono}, T., {et~al.} 2011, \apjl, 733, L39

\bibitem[{{Wyatt}(2005)}]{Wyatt2005}
{Wyatt}, M.~C. 2005, \aap, 433, 1007

\bibitem[{{Wyatt} {et~al.}(2007){Wyatt}, {Smith}, {Su}, {Rieke}, {Greaves},
  {Beichman}, \& {Bryden}}]{Wyatt2007}
{Wyatt}, M.~C., {Smith}, R., {Su}, K.~Y.~L., {et~al.} 2007, \apj, 663, 365

\bibitem[{{Yi} {et~al.}(2001){Yi}, {Demarque}, {Kim}, {Lee}, {Ree}, {Lejeune},
  \& {Barnes}}]{Yi2001}
{Yi}, S., {Demarque}, P., {Kim}, Y.-C., {et~al.} 2001, \apjs, 136, 417

\bibitem[{{Zuckerman} \& {Song}(2004)}]{ZuckermanSong2004}
{Zuckerman}, B. \& {Song}, I. 2004, \araa, 42, 685

\end{thebibliography}

\begin{appendix}
\section{HD 172555 photometry compilation}
Table \ref{Phot} shows a compilation of current literature and
Herschel photometry for \object{HD~172555}.
\section{How to derive the oxygen mass}\label{OxMass}
In the following we explain how to derive the atomic oxygen gas mass
from prompt emission level population. To estimate the mass of oxygen
gas we consider the excitation of atomic oxygen to its first fine
structure level in the absence of a collisional partner. This
situation happens in very low density environments such as debris
discs. The main mechanism involved is the so-called prompt emission
and fluorescence. The prompt emission involves the absorption of a
photon from the star or from the dust at the precise wavelength of the
atomic emission, and subsequent re-emission. To model the emission, we
assume that the ground state is the most populated. The population at
steady-state for level $\rm^{3} P_{2}$ is given by:

\begin{equation}
\centering
0=-n_1(A_{10}+B_{10}J_{10})+n_0B_{01}J_{10}.
\end{equation}
Assuming that only the first two levels are populated
($n=n_{0}+n_{1}$),
\begin{equation}
\centering
x=\frac{n_1}{n}=\frac{1}{1+B_{10}/B_{01}+A_{10}/B_{01}J_{10}}
\end{equation}
since $g_0B_{01}=g_1B_{10}$ and $B_{10}=(c^2/2h\nu^3)A_{10}$, the
fractional population is
\begin{equation}
x=\frac{n_1}{n}=\frac{1}{1+g_0/g_1+(g_0/g_1)(2h\nu^3/c^2J_{10})}.
\end{equation}
For OI we have $g_0=5$, $g_1=3$. $\rm J_{10}$ is computed using the
distance-dilluted stellar flux in the Rayleigh-Jeans regime at 63 $\rm
\mu m$:
\begin{equation}
J_{10}=\frac{1}{4\pi}\left(\frac{R_*}{R}\right)^2\frac{2kT_*}{\lambda^2},
\end{equation}
where $R$ is the distance to the star and $R_*$ the stellar radius.
We obtain $\rm J_{10}= 2.273 \times 10^{-13} \times (1AU/R(AU))^{2}$
erg s$^{-1}$ cm$^{-2}$ Hz$^{-1}$ sr$^{-1}$, and the fractional
population is thus: $x$ = $\rm 8.67 \times 10^{-5}$

The number of oxygen atoms given the flux in cgs is then
\begin{equation}
{\rm N_{[OI]}}= 9.5216\times 10^{36} \frac{F_{\rm line}(erg\ s^{-1}
  \ cm^{-2})\times D^2(pc^2)}{\frac{h\nu}{4\pi}A_{ul}\ x} R(AU)^{2},
\end{equation}
where $9.5216\times 10^{36}$ combines the conversion from $\rm
arcsec^{2}$ to steradian and from AU to parsec.

The effect of including additional excitation paths is negligible.  We
compare this with a situation when the emission is produced at 1 AU from
the star. For example, if we include all main fluorescence pumping
paths ($^2P_2$ $\rightarrow$ $^1S_0 $ $\rightarrow$ $^3P_1$, $^3P_2$
$\rightarrow$ $^1D_2$ $\rightarrow$ $^3P_1$, and $^3P_2$ $\rightarrow$
$^1S_0$ $\rightarrow$ $^1D_2$ $\rightarrow$ $^3P_1$), the final mass of
oxygen is $\rm M([OI]) =$ 0.0247 $\rm M_\oplus$, compared to 
$\rm M([OI]) =$ 0.025 $\rm M_\oplus$ obtained when we only include 
the main excitation path.

\scriptsize{
\begin{table}
\caption{Available photometry}             
\label{Phot}      
\centering          
\begin{tabular}{llllllc}     
\hline\hline       
Filter & $\rm \lambda_{0}$ & mag & $\rm \sigma_{mag}$ & flux & $\rm\sigma_{flux}$& Ref \\ 
-- & ($\rm \mu m$) & (mag) & (mag) & (Jy) & (Jy) \\ 
\hline 
Stromgren u & 0.35 & 6.354 & 0.010 & 13.69 & 0.13 & 6  \\
Johnson U & 0.37 & 5.070 & 0.025 & 17.63 & 0.40 & 1  \\
Stromgren v & 0.41 & 5.203 & 0.009 & 40.49 & 0.34 & 6 \\
Tycho B & 0.42 & 5.015 & 0.014 & 38.89 & 0.50 & 3 \\
Johnson B & 0.44 & 4.790 & 0.019 & 53.87 & 0.94 & 1 \\
Stromgren b & 0.47 & 4.891 & 0.005 & 47.21 & 0.22 & 6 \\
Tycho V & 0.53 & 4.793 & 0.009 & 45.51 & 0.38 & 3  \\
Johnson V & 0.55 & 4.990 & 0.015 & 38.45 & 0.53 & 1  \\
Stromgren y & 0.55 & 4.779 & 0.003 & 45.40 & 0.13 & 6 \\
Johnson R & 0.70 & 4.887 & 0.023 & 33.40 & 0.71 & 1  \\
Johnson I & 0.90 & 4.581 & 0.025 & 35.74 & 0.82  & 1 \\
2MASS J & 1.25 & 4.382 & 0.26 & 28.92 & 6.93 & 2  \\
2MASS H & 1.65 & 4.251 & 0.212 & 22.24 & 4.34 & 2  \\
2MASS Ks & 2.15 & 4.298 & 0.031 & 12.82 & 0.37 & 2  \\
WISE1 & 3.35 & 4.269 & 0.094 & 6.069 & 0.52 & 11 \\
IRAC1 & 3.56 & -- & -- & 4.949 & 0.035 & 13  \\
IRAC2 & 4.51 & -- & -- & 3.238 & 0.038 & 13 \\
WISE2 & 4.60 & 3.775 & 0.059 & 5.309 & 0.29 & 11 \\
IRAC3 & 5.76 & -- & -- & 2.214 & 0.007 & 13 \\
IRAC4 & 7.96 & -- & -- & 1.458 & 0.004 & 13 \\
AKARI9 & 9 & -- & -- & 1.451 & 0.017 & 10  \\
WISE3 & 11.56 & 3.615 & 0.015 & 1.134 & 0.016 & 11 \\
Si-5 & 11.7 & -- & -- & 1.155 & 0.002 & 8  \\
WISE4 & 22.09 & 2.348 & 0.020 & 0.962 & 0.018 & 11  \\
AKARI18 & 18 & -- & -- & 0.921 & 0.020 & 10  \\
Qa & 18.3 & -- & -- & 1.094 & 0.011 & 8 \\
MIPS24 & 23.84 & -- & -- & 0.766 & 0.030 & 5 \\
MIPS70 & 71.42 & -- & -- & 0.197 & 0.020 & 5 \\
PACS70 & 70 & -- & -- & 0.191 & 0.005 & 7 \\
PACS100 & 100 & -- & -- & 0.089 & 0.003 & 7 \\
PACS160 & 160 & -- & -- & 0.036 & 0.002 & 7 \\
LABOCA & 870 & -- & -- & 0.040 & 0.01 & 9 \\
SIMBA & 1200 & -- & -- & $\rm <$ 0.026 & 0. & 12 \\
\hline                  
\end{tabular}
\tablefoot{(1): \cite{Johson1966}; (2): 2MASS catalogue
  \citep{Skrutskie2006}; (3): Tycho-2 catalogue \citep{Hog2000}; (4)
  IRAS faint source catalogue; (5): \cite{Rebull2008}; (6)
  \cite{Hauck1998}; (7): present work; (8): \cite{Moerchen2010}; (9):
  \cite{Nilsson2009}, see text; (10): \cite{Ishihara2010}; (11): WISE
  Preliminary Release Source Catalog. The WISE magnitude limits
in the Explanatory Suplement indicate that the 3.4, 4.6 and 12 $\rm
\mu m$ photometry of \object{HD~172555} is saturated. However, the
WISE pipeline performs PSF fitting when dealing with saturated sources
and produces accurate results for bands 1, 3, and 4. Unfortunately, for
band 2 at 4.6 $\rm \mu m$ the pipeline does not produce good results
for saturated sources (K. Stapelfeldt, private communication).;
(12): \cite{Schutz2005}; (13): photometry  reduced by us with MOPEX
from IRAC archive data. IRAC1, IRAC2 and IRAC3 measurements are saturated.}
\end{table}
}
\end{appendix}
\end{document}